\documentclass[11pt,preprint]{aastex}

\usepackage{longtable}
\newcommand{\etal}{{\it et~al.}}

\begin{document}

\title{A revised asteroid polarization-albedo relationship using WISE/NEOWISE data}

\author{Joseph R. Masiero\altaffilmark{1}, A. K. Mainzer\altaffilmark{1}, T. Grav\altaffilmark{2}, J. M. Bauer\altaffilmark{1,3}, E. L. Wright\altaffilmark{4}, R. S. McMillan\altaffilmark{5}, D. J. Tholen\altaffilmark{6}, A. W. Blain\altaffilmark{7}}

\altaffiltext{1}{Jet Propulsion Laboratory/California Institute of Technology, 4800 Oak Grove Dr., MS 321-520, Pasadena, CA 91109, USA, {\it Joseph.Masiero@jpl.nasa.gov}}
\altaffiltext{2}{Planetary Science Institute, 1700 East Fort Lowell, Suite 106, Tucson, AZ 85719-2395}
\altaffiltext{3}{Infrared Processing and Analysis Center, California Institute of Technology, Pasadena, CA 91125 USA}
\altaffiltext{4}{UCLA Astronomy, PO Box 91547, Los Angeles, CA 90095-1547 USA}
\altaffiltext{5}{Lunar and Planetary Laboratory, University of Arizona, 1629 East University Blvd, Kuiper Space Science Bldg. \#92, Tucson, AZ 85721-0092 USA}
\altaffiltext{6}{Institute for Astronomy, University of Hawaii, Honolulu, HI 96822 USA}
\altaffiltext{7}{Department of Physics and Astronomy, University of Leicester, University Road, Leicester, LE1 7RH, United Kingdom}

\begin{abstract}

We present a reanalysis of the relationship between asteroid albedo
and polarization properties using the albedos derived from the
Wide-field Infrared Survey Explorer.  We find that the function that
best describes this relation is a three-dimensional linear fit in the
space of $\log$ (albedo)-$\log$ (polarization slope)-$\log$ (minimum
polarization).  When projected to two dimensions the parameters of the
fit are consistent with those found in previous work.  We also define
$p^\star$ as the quantity of maximal polarization variation when
compared with albedo and present the best fitting albedo-$p^\star$
relation.  Some asteroid taxonomic types stand out in this
three-dimensional space, notably the E, B, and M Tholen types, while
others cluster in clumps coincident with the S- and C-complex bodies.
We note that both low albedo and small ($D<30~$km) asteroids are
under-represented in the polarimetric sample, and we encourage future
polarimetric surveys to focus on these bodies.

\end{abstract}

\section{Introduction}

As light scatters off the surface of atmosphereless bodies, it is
instilled with a small linear polarization.  The degree of linear
polarization of the scattered light measured by the observer is a
function of the phase angle of observation and the composition and
structure of the surface, in particular the interrelated parameters of
albedo, index of refraction, and space between scattering elements
\citep[e.g.][]{muinonen89,shkuratov94}.  Early work quantified the
relation between phase angle (the angle between the direction to the
sun and the observer as seen from the target, $\alpha$) and
polarization \citep{dollfus79} and this effect can be used in parallel
with the magnitude-phase effect to probe the scattering physics of
atmosphereless surfaces \citep{muinonen02,muinonen09}.

As expected from classical scattering models, the light reflected from
a surface is polarized perpendicular to the scattering plane for large
phase angles, which is referred to as a positive polarization.  For
small phase angles, however, light acquires a polarization in the
scattering plane due to an increase in the dominance of second-order
scattering.  This case is referred to as negative polarization, as it
is perpendicular to the positive case and thus carries a negative sign
when the polarization coordinate system is rotated to account for the
viewing geometry.  The angle where the phase curve transitions from
positive to negative is referred to as the inversion angle
($\alpha_0$).  By definition the value of the polarization must go to
zero at $\alpha=0^\circ$, though some work has suggested that surfaces
may have a secondary trough at very small angles related to the
optical opposition effect \citep{rosenbush97}.  \citet{cellino05a}
find no evidence for a polarimetric opposition effect in their sample,
though high albedo objects are not represented there.

\begin{figure}[ht]
\begin{center}
\includegraphics[scale=0.6]{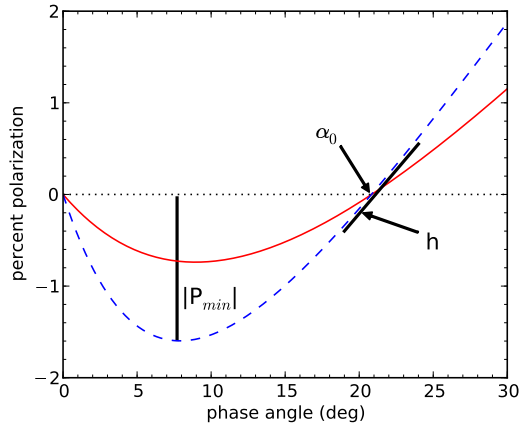} 
\protect\caption{Illustration of the typical polarization-phase behavior for two different types of asteroids, with the inversion angle ($\alpha_0$), the minimum polarization ($P_{min}$), and the slope of polarization at the inversion angle ($h$) labeled for the dashed blue curve.  Example curves for a generic S-type asteroid (solid red line) and a generic C-type (dashed blue line) are shown.}
\label{fig.phasefig}
\end{center}
\end{figure}

From the studies of the scattering properties of the lunar surface, a
relationship was found between albedo and the parameters used to
describe the polarimetric-phase effect of the lunar regolith
\citep{bowell73} that was then extended to asteroids
\citep{zellner74,cellino99}, of the form:
\begin{eqnarray}
\log p_V = C_1 \log h + C_2 \label{eq.slope} \\
\log p_V = C_3 \log P_{min} + C_4 \label{eq.pmin}
\end{eqnarray}
where $p_V$ is the geometric albedo, $h$ is the linear slope of the
phase curve at the inversion angle, and $P_{min}$ is the value of the
largest negative polarization (i.e. the depth of the negative trough),
usually expressed as an absolute value.  We show an illustration of
these parameters and two typical polarization-phase curves in
Figure~\ref{fig.phasefig}.  We note that the polarization shown in
this figure is the $P_r$ value that has been rotated to account for
viewing geometry, such that $P_r>0$ is the amplitude of polarization
perpendicular to the scattering plane and $P_r<0$ is the amplitude
parallel to the scattering plane.  A polarization component $\pm
45^\circ$ from the scattering plane is typically not observed at any
phase angle for asteroids, and thus is ignored in this diagram.

\citet{cellino99} present the most recent best-fitting values for the
constants in the above equations: $C_1=-1.118\pm0.071$,
$C_2=-1.779\pm0.062$, $C_3=-1.357\pm0.140$, and $C_4=-0.858\pm0.030$.
In this work, we revise the best-fitting values for these constants in
light of new albedo data from the Wide-field Infrared Survey Explorer
\citep[WISE,][]{wright10} and the planetary science extension NEOWISE
\citep{mainzer11}.  The aim of this work is two-fold: firstly, while
WISE provides us with albedos for a large fraction of the known
asteroids, calibration of this relationship will allow it to be
applied to objects that were not observed by WISE; secondly, the
behavior of the polarization of asteroid surfaces helps us determine
the surface mineralogy and this relationship represents a critical
component in this determination.  Through application of both thermal
infrared and polarimetric data we can gain a broader understanding of
the behavior of asteroids across the Solar system.

\section{Data}

We draw our list of polarimetric properties for asteroids from a range
of sources.  The dominant contributor is the Astronomical Polarimetric
Database presented in the Planetary Data System (PDS)
\citep{lupishko08} which was a compilation of the polarimetric
properties of individual asteroids in the literature up to the data of
publication.  We also incorporate values for $h$, $P_{min}$, and/or
$\alpha_0$ for asteroids given by: \citet{cellino99, cellino05a,
  cellino05b, fornasier06, gilhutton07a, gilhutton07b, gilhutton08,
  masiero09, belskaya10}.  We note that as these data are drawn from a
range of different instruments, uncertainties in the absolute
calibration may result in a larger scatter than is actually present.
A comprehensive survey of polarimetric properties of a large number of
objects conducted with a single instrument would reduce this possible
source of error, and so is strongly encouraged.

Determination of $h$, $P_{min}$, and $\alpha_0$ all require
polarimetric measurements spanning a range of phase angles.  The
inversion angle can typically be determined to a reasonable level of
accuracy with a few bounding measurements at $\alpha\sim20^\circ$.
Polarimetric slope is more difficult to determine, especially for
objects located farther from the sun that are rarely observable at
phase angles much beyond the inversion angle (e.g. objects that do not
come within $\sim2.9~$AU of the sun can never be observed at phase
angles $\alpha>20^\circ$).  Careful timing of observations can ensure
adequate phase coverage that will allow for an accurate determination
of the slope.  The depth of minimum polarization is often the most
difficult parameter to determine for some objects, as it requires
observing at small phase angles that are not frequently available for
asteroids in the inner Main Belt, Hungaria, Mars Crosser, and NEO
populations.  Additionally, determining this value requires evenly
spaced observations over the full branch of negative polarization,
rather than just a few bounding measurements as required for both
$\alpha_0$ and $h$.  As such, relative errors on $P_{min}$ tend to be
larger than measured for the other polarimetric parameters.  Where
errors on polarimetric parameters were not given by the source, we
assume values based on the errors on the published data in those
sources.

The albedos we use for this work are drawn from the values derived for
Main Belt asteroids (MBAs) published in \citet{masiero11}.  For
objects in the NEO or Mars Crossing populations, we draw albedos from
the appropriate lists \citep{mainzer11neo,mainzer11mc} which were
derived using a method identical to that used for the MBAs.  All of
the objects with both defined polarimetric phase curves as well as
WISE-determined albedos had low identifying numbers, implying that
they were some of the first objects discovered, and thus likely to
preferentially sample the largest minor bodies of the solar system.
These large asteroids were more likely to have been seen in multiple
bands by WISE which allows for fitting of the beaming parameter.
\citet{mainzer11cal} show that in cases such as this the error on
albedo as an absolute measurement is $\sim20\%$ of the measured albedo
value, however internal comparisons are better than this limit.

A primary concern in any analysis of albedos derived from
infrared-determined diameters is the quality of the optical
measurements used.  We draw our $H$ magnitudes from the Minor Planet
Center's orbital element catalog (MPCORB\footnote{{\it
    http://www.minorplanetcenter.net/iau/MPCORB.html}}), as discussed
in \citet{masiero11}.  While other studies have found an offset
between measured magnitudes and those predicted from the $H$ absolute
magnitude value, with objects being fainter than expected
\citep[e.g.][]{juric02,parker08}, \citet{mainzer11cal} find that, in
general, no offset corrections to $H$ are required for the most recent
releases of the magnitudes.  An exception to this result has been
found for some objects with unusually high albedos in
\citet{masiero11} and \citet{mainzer11neo}; many of these objects are
coincident in orbital-element-space with the Hungarias and the Vesta
family.  \citet{harris89} found that the commonly assumed value of
$G=0.15$ is inappropriate for these types of high-albedo objects, and
a value of $G\sim0.4$ may be more appropriate.  Revising $G$ to this
value would result in an offset of up to $\sim0.3~$mag in the $H$
magnitude depending on the initial $H$ fit, however this correction is
not required for most objects.  In the past, photometric measurements
for many asteroids that contributed to the $H$ magnitudes in the
MPCORB catalog were acquired with unfiltered CCDs.  New, filtered
observations and refined handling of previous photometry have largely
mitigated the effect of unfiltered measurements on $H$ values.
(T. Spahr, 2012, private communication).  

\citet{mainzer11iras} present a comparison of the WISE albedos to the
IRAS albedos and find a good match for most objects, though some
scatter is seen, especially at the smallest sizes where the IRAS
signal-to-noise ratios were poor compared to WISE.  This albedo error
assumes moderate-to-low light curve amplitudes and well characterized
$H$ and $G$ values.  This error will result in an uncertainty in the
offsets of the linear fits (i.e. $C_2$ and $C_4$ in
Equations~\ref{eq.slope} and \ref{eq.pmin}) though the slopes should
be unaffected.  We include this error in our fits below.  We note that
recently \citet{muinonen10} have introduced a three-parameter
photometric system ($H$,$G_1$,$G_2$) to better characterize the
behavior of the photometric phase effect which may reduce some of
these errors, but we note that this system requires accurate
photometry over a large phase window, which is not available for many
asteroids.

We show the polarimetric and albedo data used for this work in Table
\ref{tab.data} (ellipses indicate an unmeasured polarimetric
property).  When these two data sets are combined, we have $65$
objects with measured albedos and polarimetric slopes, and $112$ with
albedos and minimum polarization values.  This result is a improvement
compared to the data set presented by \citet{cellino99}, who performed
a similar analysis using IRAS albedos of $37$ objects for the
slope-albedo fit and $16$ for the minimum polarization-albedo fit.  We
note that due to the brightness requirements of most polarimeters and
the polarimetric survey strategies employed, these lists are dominated
by the largest known asteroids.  Approximately half our sample have
sizes over $100~$km, and three-quarters are larger than $50~$km.  Thus
while the largest asteroids are well sampled, there is a distinct lack
of small bodies in these lists.  We also note that despite the fact
that low-albedo objects dominate the Main Belt population
\citep{masiero11}, they are under-represented in the polarimetric
surveys (see below).  As WISE is sensitive to thermal infrared light,
the detection probability for asteroids is effectively unbiased with
respect to the albedos of the objects observed \citep{mainzer11neo},
and thus the distribution of albedos seen with WISE is a more accurate
representation of the true population than is the distribution seen
for optically-selected samples.  Extending polarimetric coverage to
both smaller sizes and low albedo objects through a large-scale
campaign is critical to extending and generalizing the trends seen
here and in previous work.

\begin{center}
\begin{longtable}{ccccc}
\caption{Compiled asteroid albedos and polarimetric properties}\label{tab.data}\\
\hline\hline
Asteroid &       $p_V$       &   $\alpha_0$  &     $h$     &    $P_{min}$    \\ 
\hline\hline
\endfirsthead
\caption[]{(continued)}\\
\hline\hline
Asteroid &       $p_V$       &   $\alpha_0$  &     $h$     &    $P_{min}$    \\ 
\hline\hline
\endhead
\hline
\endfoot
    2  & $0.142 \pm 0.018$ & $18.1 \pm 0.1$ & $0.228 \pm 0.003$ & $1.38 \pm 0.05$ \\ 
    5  & $0.245 \pm 0.051$ & $19.1 \pm 0.1$ & $0.096 \pm 0.050$ & $0.70 \pm 0.05$ \\ 
    6  & $0.269 \pm 0.049$ & $20.8 \pm 0.2$ & $0.091 \pm 0.050$ & $0.80 \pm 0.05$ \\ 
    8  & $0.261 \pm 0.048$ & $20.0 \pm 0.1$ & $0.104 \pm 0.003$ & $0.68 \pm 0.05$ \\ 
    9  & $0.134 \pm 0.016$ & $21.8 \pm 0.1$ & $0.102 \pm 0.003$ & $0.74 \pm 0.05$ \\ 
   10  & $0.058 \pm 0.005$ & $     ...    $ & $      ...      $ & $1.50 \pm 0.05$ \\ 
   11  & $0.158 \pm 0.036$ & $18.9 \pm 0.2$ & $0.124 \pm 0.003$ & $0.73 \pm 0.05$ \\ 
   12  & $0.140 \pm 0.014$ & $20.8 \pm 0.2$ & $0.121 \pm 0.003$ & $0.73 \pm 0.01$ \\ 
   13  & $0.069 \pm 0.022$ & $21.7 \pm 0.5$ & $0.257 \pm 0.003$ & $2.10 \pm 0.05$ \\ 
   14  & $0.221 \pm 0.022$ & $20.5 \pm 0.2$ & $0.105 \pm 0.003$ & $0.82 \pm 0.10$ \\ 
   15  & $0.206 \pm 0.055$ & $20.6 \pm 0.2$ & $0.087 \pm 0.005$ & $0.72 \pm 0.02$ \\ 
   17  & $0.160 \pm 0.009$ & $     ...    $ & $0.131 \pm 0.003$ & $0.74 \pm 0.05$ \\ 
   18  & $0.221 \pm 0.082$ & $21.6 \pm 0.1$ & $0.101 \pm 0.003$ & $0.87 \pm 0.05$ \\ 
   19  & $0.050 \pm 0.020$ & $21.7 \pm 0.2$ & $0.305 \pm 0.003$ & $1.72 \pm 0.05$ \\ 
   22  & $0.169 \pm 0.061$ & $     ...    $ & $      ...      $ & $0.83 \pm 0.04$ \\ 
   24  & $0.064 \pm 0.016$ & $     ...    $ & $0.191 \pm 0.003$ & $1.63 \pm 0.10$ \\ 
   27  & $0.201 \pm 0.058$ & $     ...    $ & $0.099 \pm 0.003$ & $0.70 \pm 0.05$ \\ 
   29  & $0.157 \pm 0.035$ & $22.0 \pm 0.2$ & $0.098 \pm 0.003$ & $0.88 \pm 0.10$ \\ 
   30  & $0.171 \pm 0.034$ & $19.8 \pm 0.5$ & $0.104 \pm 0.003$ & $0.78 \pm 0.05$ \\ 
   31  & $0.045 \pm 0.044$ & $     ...    $ & $      ...      $ & $1.32 \pm 0.10$ \\ 
   32  & $0.230 \pm 0.065$ & $     ...    $ & $      ...      $ & $0.63 \pm 0.05$ \\ 
   39  & $0.245 \pm 0.056$ & $21.0 \pm 0.1$ & $0.090 \pm 0.003$ & $0.79 \pm 0.05$ \\ 
   40  & $0.195 \pm 0.019$ & $20.8 \pm 0.2$ & $0.100 \pm 0.003$ & $0.85 \pm 0.05$ \\ 
   46  & $0.052 \pm 0.011$ & $     ...    $ & $      ...      $ & $1.54 \pm 0.10$ \\ 
   47  & $0.067 \pm 0.009$ & $     ...    $ & $0.204 \pm 0.003$ & $1.44 \pm 0.05$ \\ 
   51  & $0.100 \pm 0.026$ & $     ...    $ & $0.292 \pm 0.050$ & $1.86 \pm 0.05$ \\ 
   54  & $0.049 \pm 0.008$ & $22.2 \pm 0.5$ & $0.357 \pm 0.050$ & $1.95 \pm 0.05$ \\ 
   56  & $0.050 \pm 0.006$ & $19.7 \pm 0.2$ & $0.318 \pm 0.003$ & $1.47 \pm 0.05$ \\ 
   57  & $0.182 \pm 0.047$ & $     ...    $ & $      ...      $ & $0.71 \pm 0.10$ \\ 
   58  & $0.059 \pm 0.005$ & $     ...    $ & $      ...      $ & $1.70 \pm 0.10$ \\ 
   63  & $0.159 \pm 0.028$ & $19.8 \pm 0.1$ & $0.102 \pm 0.003$ & $0.70 \pm 0.05$ \\ 
   64  & $0.676 \pm 0.223$ & $18.2 \pm 0.2$ & $0.036 \pm 0.001$ & $0.32 \pm 0.05$ \\ 
   68  & $0.207 \pm 0.025$ & $     ...    $ & $      ...      $ & $0.68 \pm 0.05$ \\ 
   70  & $0.040 \pm 0.009$ & $     ...    $ & $      ...      $ & $1.83 \pm 0.05$ \\ 
   71  & $0.248 \pm 0.035$ & $     ...    $ & $0.061 \pm 0.005$ & $0.61 \pm 0.02$ \\ 
   73  & $0.186 \pm 0.018$ & $     ...    $ & $      ...      $ & $0.76 \pm 0.06$ \\ 
   75  & $0.098 \pm 0.014$ & $20.2 \pm 0.1$ & $0.103 \pm 0.017$ & $     ...     $ \\ 
   77  & $0.153 \pm 0.046$ & $     ...    $ & $      ...      $ & $1.25 \pm 0.14$ \\ 
   80  & $0.182 \pm 0.026$ & $     ...    $ & $      ...      $ & $0.75 \pm 0.05$ \\ 
   83  & $0.086 \pm 0.021$ & $     ...    $ & $      ...      $ & $1.47 \pm 0.10$ \\ 
   84  & $0.053 \pm 0.017$ & $20.3 \pm 0.5$ & $0.306 \pm 0.050$ & $1.49 \pm 0.05$ \\ 
   85  & $0.063 \pm 0.025$ & $     ...    $ & $      ...      $ & $1.36 \pm 0.10$ \\ 
   89  & $0.185 \pm 0.034$ & $     ...    $ & $0.119 \pm 0.050$ & $0.90 \pm 0.05$ \\ 
   95  & $0.056 \pm 0.009$ & $     ...    $ & $      ...      $ & $1.78 \pm 0.05$ \\ 
   97  & $0.206 \pm 0.046$ & $22.1 \pm 0.1$ & $0.174 \pm 0.018$ & $     ...     $ \\ 
  113  & $0.223 \pm 0.031$ & $     ...    $ & $0.081 \pm 0.005$ & $     ...     $ \\ 
  114  & $0.088 \pm 0.010$ & $     ...    $ & $      ...      $ & $1.24 \pm 0.10$ \\ 
  115  & $0.654 \pm 0.124$ & $     ...    $ & $      ...      $ & $0.71 \pm 0.05$ \\ 
  118  & $0.139 \pm 0.031$ & $     ...    $ & $      ...      $ & $0.80 \pm 0.12$ \\ 
  121  & $0.077 \pm 0.010$ & $     ...    $ & $      ...      $ & $1.72 \pm 0.05$ \\ 
  125  & $0.115 \pm 0.027$ & $     ...    $ & $0.145 \pm 0.033$ & $0.83 \pm 0.02$ \\ 
  129  & $0.157 \pm 0.026$ & $     ...    $ & $      ...      $ & $0.90 \pm 0.05$ \\ 
  131  & $0.164 \pm 0.011$ & $     ...    $ & $0.208 \pm 0.059$ & $     ...     $ \\ 
  132  & $0.120 \pm 0.008$ & $     ...    $ & $0.146 \pm 0.006$ & $1.13 \pm 0.03$ \\ 
  135  & $0.152 \pm 0.050$ & $     ...    $ & $      ...      $ & $1.06 \pm 0.10$ \\ 
  138  & $0.161 \pm 0.028$ & $     ...    $ & $0.103 \pm 0.020$ & $     ...     $ \\ 
  139  & $0.045 \pm 0.023$ & $     ...    $ & $0.262 \pm 0.050$ & $1.31 \pm 0.05$ \\ 
  141  & $0.049 \pm 0.010$ & $20.6 \pm 0.5$ & $0.330 \pm 0.050$ & $1.78 \pm 0.05$ \\ 
  145  & $0.043 \pm 0.004$ & $     ...    $ & $      ...      $ & $1.86 \pm 0.05$ \\ 
  153  & $0.046 \pm 0.008$ & $     ...    $ & $      ...      $ & $1.05 \pm 0.05$ \\ 
  182  & $0.210 \pm 0.059$ & $     ...    $ & $      ...      $ & $0.64 \pm 0.05$ \\ 
  184  & $0.107 \pm 0.019$ & $     ...    $ & $      ...      $ & $0.93 \pm 0.06$ \\ 
  188  & $0.157 \pm 0.055$ & $     ...    $ & $0.140 \pm 0.015$ & $     ...     $ \\ 
  189  & $0.199 \pm 0.024$ & $     ...    $ & $      ...      $ & $1.26 \pm 0.10$ \\ 
  192  & $0.288 \pm 0.040$ & $19.8 \pm 0.1$ & $0.084 \pm 0.003$ & $0.75 \pm 0.05$ \\ 
  197  & $0.239 \pm 0.026$ & $     ...    $ & $      ...      $ & $0.79 \pm 0.08$ \\ 
  201  & $0.097 \pm 0.006$ & $     ...    $ & $      ...      $ & $1.00 \pm 0.05$ \\ 
  204  & $0.163 \pm 0.044$ & $     ...    $ & $      ...      $ & $0.83 \pm 0.12$ \\ 
  216  & $0.111 \pm 0.034$ & $     ...    $ & $      ...      $ & $1.27 \pm 0.05$ \\ 
  217  & $0.044 \pm 0.005$ & $     ...    $ & $      ...      $ & $0.82 \pm 0.05$ \\ 
  230  & $0.171 \pm 0.076$ & $20.6 \pm 0.2$ & $0.122 \pm 0.003$ & $0.94 \pm 0.05$ \\ 
  234  & $0.151 \pm 0.034$ & $27.0 \pm 2.0$ & $      ...      $ & $1.60 \pm 0.20$ \\ 
  250  & $0.112 \pm 0.021$ & $     ...    $ & $      ...      $ & $0.88 \pm 0.08$ \\ 
  259  & $0.042 \pm 0.005$ & $     ...    $ & $      ...      $ & $1.25 \pm 0.05$ \\ 
  270  & $0.254 \pm 0.043$ & $     ...    $ & $      ...      $ & $0.65 \pm 0.05$ \\ 
  305  & $0.182 \pm 0.028$ & $     ...    $ & $      ...      $ & $0.64 \pm 0.10$ \\ 
  306  & $0.174 \pm 0.060$ & $     ...    $ & $      ...      $ & $0.66 \pm 0.10$ \\ 
  324  & $0.063 \pm 0.012$ & $20.0 \pm 0.1$ & $0.278 \pm 0.003$ & $1.46 \pm 0.05$ \\ 
  334  & $0.051 \pm 0.016$ & $     ...    $ & $      ...      $ & $1.32 \pm 0.05$ \\ 
  338  & $0.165 \pm 0.028$ & $     ...    $ & $      ...      $ & $0.98 \pm 0.10$ \\ 
  345  & $0.059 \pm 0.012$ & $     ...    $ & $      ...      $ & $1.55 \pm 0.05$ \\ 
  347  & $0.213 \pm 0.041$ & $22.6 \pm 0.1$ & $0.113 \pm 0.003$ & $0.78 \pm 0.03$ \\ 
  349  & $0.153 \pm 0.018$ & $     ...    $ & $      ...      $ & $0.39 \pm 0.05$ \\ 
  351  & $0.171 \pm 0.046$ & $     ...    $ & $      ...      $ & $0.74 \pm 0.09$ \\ 
  354  & $0.173 \pm 0.032$ & $     ...    $ & $      ...      $ & $0.51 \pm 0.10$ \\ 
  356  & $0.053 \pm 0.015$ & $     ...    $ & $      ...      $ & $1.50 \pm 0.10$ \\ 
  377  & $0.056 \pm 0.025$ & $19.8 \pm 0.2$ & $0.206 \pm 0.005$ & $1.76 \pm 0.04$ \\ 
  384  & $0.190 \pm 0.040$ & $     ...    $ & $      ...      $ & $0.94 \pm 0.35$ \\ 
  396  & $0.139 \pm 0.025$ & $     ...    $ & $      ...      $ & $1.34 \pm 0.09$ \\ 
  409  & $0.050 \pm 0.010$ & $19.9 \pm 0.2$ & $0.191 \pm 0.005$ & $     ...     $ \\ 
  410  & $0.043 \pm 0.007$ & $     ...    $ & $0.313 \pm 0.050$ & $1.94 \pm 0.05$ \\ 
  415  & $0.086 \pm 0.009$ & $     ...    $ & $      ...      $ & $1.28 \pm 0.10$ \\ 
  423  & $0.066 \pm 0.005$ & $     ...    $ & $      ...      $ & $1.40 \pm 0.05$ \\ 
  441  & $0.139 \pm 0.026$ & $     ...    $ & $      ...      $ & $1.41 \pm 0.12$ \\ 
  451  & $0.069 \pm 0.006$ & $     ...    $ & $      ...      $ & $1.62 \pm 0.05$ \\ 
  466  & $0.086 \pm 0.009$ & $     ...    $ & $      ...      $ & $1.60 \pm 0.10$ \\ 
  511  & $0.073 \pm 0.006$ & $19.4 \pm 0.1$ & $0.277 \pm 0.003$ & $1.69 \pm 0.05$ \\ 
  532  & $0.202 \pm 0.039$ & $     ...    $ & $0.122 \pm 0.003$ & $0.78 \pm 0.05$ \\ 
  550  & $0.137 \pm 0.024$ & $     ...    $ & $0.157 \pm 0.005$ & $     ...     $ \\ 
  558  & $0.117 \pm 0.010$ & $     ...    $ & $      ...      $ & $0.75 \pm 0.06$ \\ 
  584  & $0.244 \pm 0.060$ & $19.1 \pm 0.1$ & $0.108 \pm 0.003$ & $0.64 \pm 0.05$ \\ 
  600  & $0.177 \pm 0.036$ & $     ...    $ & $      ...      $ & $0.43 \pm 0.20$ \\ 
  602  & $0.052 \pm 0.007$ & $     ...    $ & $      ...      $ & $1.76 \pm 0.10$ \\ 
  624  & $0.077 \pm 0.020$ & $     ...    $ & $      ...      $ & $1.30 \pm 0.05$ \\ 
  625  & $0.197 \pm 0.058$ & $     ...    $ & $0.070 \pm 0.003$ & $     ...     $ \\ 
  654  & $0.043 \pm 0.011$ & $20.5 \pm 0.5$ & $0.280 \pm 0.050$ & $1.46 \pm 0.10$ \\ 
  662  & $0.193 \pm 0.028$ & $     ...    $ & $      ...      $ & $1.32 \pm 0.33$ \\ 
  674  & $0.206 \pm 0.033$ & $     ...    $ & $      ...      $ & $0.81 \pm 0.10$ \\ 
  704  & $0.076 \pm 0.010$ & $15.7 \pm 0.1$ & $0.305 \pm 0.003$ & $1.45 \pm 0.10$ \\ 
  737  & $0.136 \pm 0.043$ & $     ...    $ & $      ...      $ & $0.84 \pm 0.05$ \\ 
  787  & $0.120 \pm 0.022$ & $     ...    $ & $0.087 \pm 0.003$ & $     ...     $ \\ 
  796  & $0.205 \pm 0.041$ & $     ...    $ & $0.124 \pm 0.011$ & $0.98 \pm 0.02$ \\ 
  849  & $0.115 \pm 0.016$ & $     ...    $ & $      ...      $ & $0.95 \pm 0.05$ \\ 
  857  & $0.225 \pm 0.026$ & $     ...    $ & $      ...      $ & $0.75 \pm 0.16$ \\ 
  863  & $0.112 \pm 0.016$ & $18.1 \pm 0.2$ & $0.052 \pm 0.005$ & $0.40 \pm 0.10$ \\ 
  887  & $0.230 \pm 0.018$ & $     ...    $ & $0.101 \pm 0.050$ & $0.76 \pm 0.05$ \\ 
  925  & $0.253 \pm 0.053$ & $19.6 \pm 0.2$ & $0.065 \pm 0.005$ & $     ...     $ \\ 
 1036  & $0.212 \pm 0.026$ & $20.6 \pm 0.2$ & $0.112 \pm 0.003$ & $0.84 \pm 0.02$ \\ 
 1052  & $0.273 \pm 0.074$ & $     ...    $ & $      ...      $ & $0.67 \pm 0.05$ \\ 
 1058  & $0.242 \pm 0.024$ & $     ...    $ & $      ...      $ & $0.69 \pm 0.10$ \\ 
 1105  & $0.102 \pm 0.017$ & $     ...    $ & $      ...      $ & $1.20 \pm 0.16$ \\ 
 1355  & $0.466 \pm 0.082$ & $18.2 \pm 0.1$ & $0.083 \pm 0.020$ & $     ...     $ \\ 
 1627  & $0.153 \pm 0.046$ & $     ...    $ & $0.131 \pm 0.003$ & $     ...     $ \\ 
 1672  & $0.094 \pm 0.016$ & $     ...    $ & $0.131 \pm 0.003$ & $     ...     $ \\ 
 1685  & $0.292 \pm 0.127$ & $     ...    $ & $0.099 \pm 0.003$ & $     ...     $ \\ 
 2131  & $0.198 \pm 0.034$ & $     ...    $ & $      ...      $ & $0.86 \pm 0.15$ \\ 
 2577  & $0.377 \pm 0.062$ & $20.9 \pm 0.1$ & $0.124 \pm 0.044$ & $     ...     $ \\ 
 3169  & $0.423 \pm 0.067$ & $19.6 \pm 0.1$ & $0.276 \pm 0.018$ & $     ...     $ \\ 
 6249  & $0.878 \pm 0.140$ & $22.4 \pm 0.1$ & $0.164 \pm 0.035$ & $     ...     $ \\ 
 6911  & $0.454 \pm 0.083$ & $     ...    $ & $      ...      $ & $0.83 \pm 0.16$ \\ 
\hline
\end{longtable}
\end{center}

\section{Revised Polarimetric-Albedo Relationship}
\label{newpol}

\begin{figure}[ht]
\begin{center}
\includegraphics[scale=0.6]{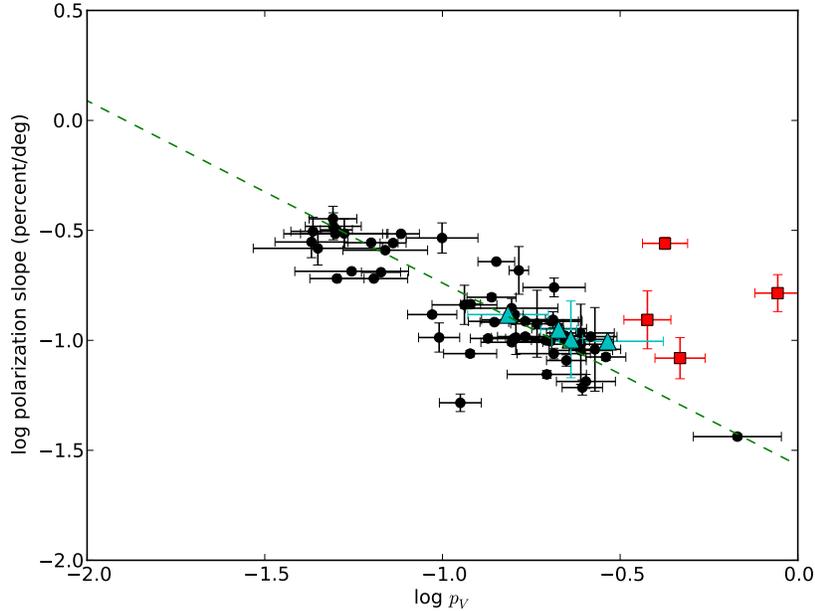} 
\protect\caption{Albedo vs. slope of the polarization phase curve beyond the inversion angle.  Objects located in the Hungaria region are noted as red squares, objects in the NEO population as cyan triangles.  The green dashed line shows the best fit found for Equation~\ref{eq.slope} with our data.}
\label{fig.ah}
\end{center}
\end{figure}

\begin{figure}[ht]
\begin{center}
\includegraphics[scale=0.6]{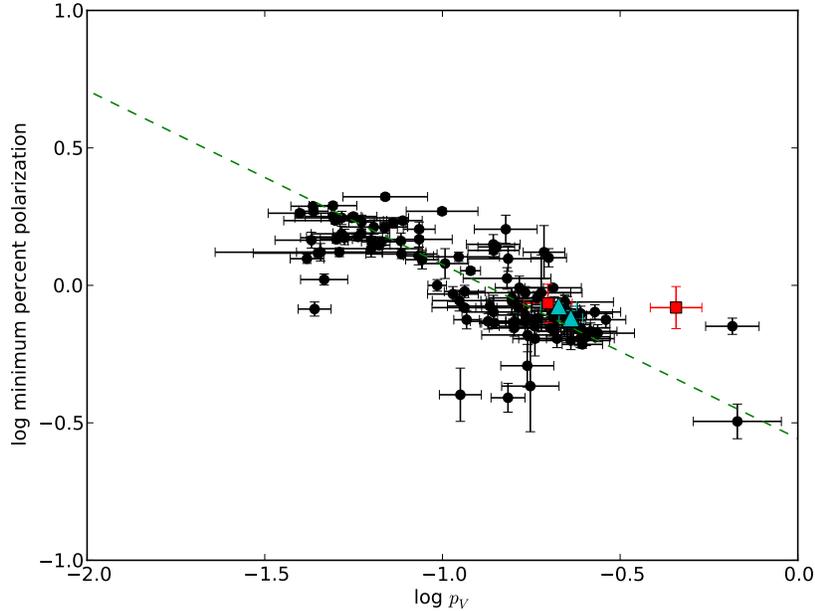} 
\protect\caption{The same as Figure~\ref{fig.ah}, but for albedo vs. depth of the minimum polarization branch.  The green dashed line shows the best fit found for Equation~\ref{eq.pmin} with our data.}
\label{fig.amin}
\end{center}
\end{figure}

\begin{figure}[ht]
\begin{center}
\includegraphics[scale=0.6]{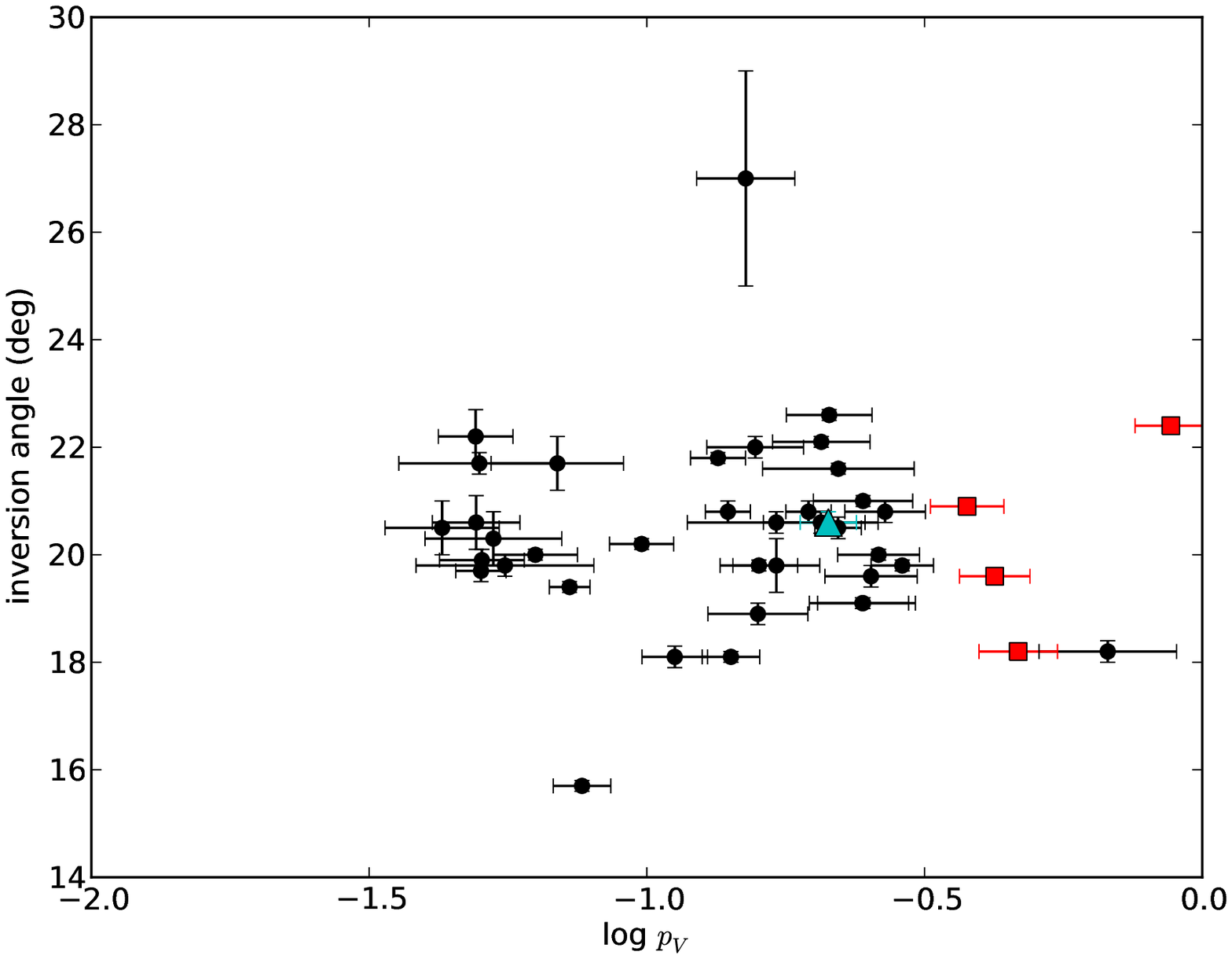} 
\protect\caption{The same as Figure~\ref{fig.ah}, but for albedo vs. inversion angle.}
\label{fig.ainv}
\end{center}
\end{figure}

In Figures~\ref{fig.ah}-\ref{fig.ainv} we compare the measured WISE
albedo to the slope of the polarization beyond the inversion angle,
the depth of the negative branch of polarization, and the inversion
angle, respectively, for all objects with recorded values for these
parameters.  We distinguish objects in the Hungaria region and in the
NEO population as red squares and cyan triangles, respectively.  While
the NEOs appear consistent with the MBAs, the Hungaria objects deviate
from the general trend significantly.  As discussed in
\citet{masiero11} the albedos for these objects are suspect: large
deviations in the magnitude-phase slope parameter from the assumed
$G=0.15$ used for most asteroids can result in incorrect $H$ values,
and thus poorly constrained albedos \citep{harris89}.  Alternatively,
large-amplitude long-period light curves may also corrupt the $H$
values calculated from optical photometry.  We are currently working
on a program to better constrain the photometric parameters and
albedos of these objects, but for the following discussion we will
disregard the Hungaria asteroids.

We see no overall trend between albedo and inversion angle in our
data.  The object with the anomalously high inversion angle is (234)
Barbara, the principal member of the ``Barbarian'' group of objects
with strange polarimetric properties \citep{cellino06}.  The object
with an inversion angle well below the general trend is (704)
Interamnia: some F-class objects like Interamnia have previously been
shown to display unusually small inversion angles \citep{belskaya05}.

We see the expected general trends when looking at slope and
P$_{min}$, with low albedo objects showing steeper slopes and deeper
troughs.  We note, however, that the lowest albedo objects, those with
$p_V<0.04$, have almost no representation in the polarimetric sample
despite representing over $10\%$ of the total population of MBAs
observed by WISE, even before correcting the population for objects
without optical followup (and thus without measured albedos).  We
therefore cannot comment on the reliability of the polarization-phase
relation at the lowest albedos.  A campaign of polarimetric
observations of low albedo asteroids is critical to test these
relations at their low albedo extreme.

We find that the optimal description of the relation between albedo
and polarimetric parameters is a linear fit in the three dimensional
space of $\log p_V$-$\log h$-$\log P_{min}$.  We use only those
objects where both polarimetric parameters are measured to an accuracy
of $20\%$ or better, leaving us with $41$ objects in our
high-confidence sample.  We perform orthogonal distance regression on
the three dimensional data, using the associated errors on each
measurement to determine the best fitting linear parameters as well as
each parameter's error.  We then reduce the best-fit parameters back
to two-dimensional projections, which result in the following constant
parameters for the relationships in Equations~\ref{eq.slope} and
\ref{eq.pmin}:
\[C_1=-1.207\pm0.067\]
\[C_2=-1.892\pm0.141\]
\[C_3=-1.579\pm0.084\]
\[C_4=-0.880\pm0.106\]
These projected fits are shown as green dashed lines in
Figures~\ref{fig.ah} and \ref{fig.amin}.  With the exception of $C_3$,
these parameters are all within one-sigma of the values found by
\citet{cellino99}, and all are within 1.5 sigma.  As the WISE albedos
for the largest asteroids have been shown to be generally consistent
with the IRAS values \citep{mainzer11iras} and all of the objects used
here are in the size range sampled by IRAS, this agreement was not
unexpected.  Of the objects in our high-confidence polarimetric sample
only $6$ were not observed by IRAS, however the WISE albedos are all
derived from a minimum of $5$ observations (and an average of $>10$)
spread over time and thus are less sensitive to rotation effects.  As
the WISE data cover MBAs down to a few kilometers and NEOs to much
smaller sizes, future polarimetric surveys focusing on smaller
asteroids will allow this relationship to be tested over a more
extensive size range.

\begin{figure}[ht]
\begin{center}
\includegraphics[scale=0.6]{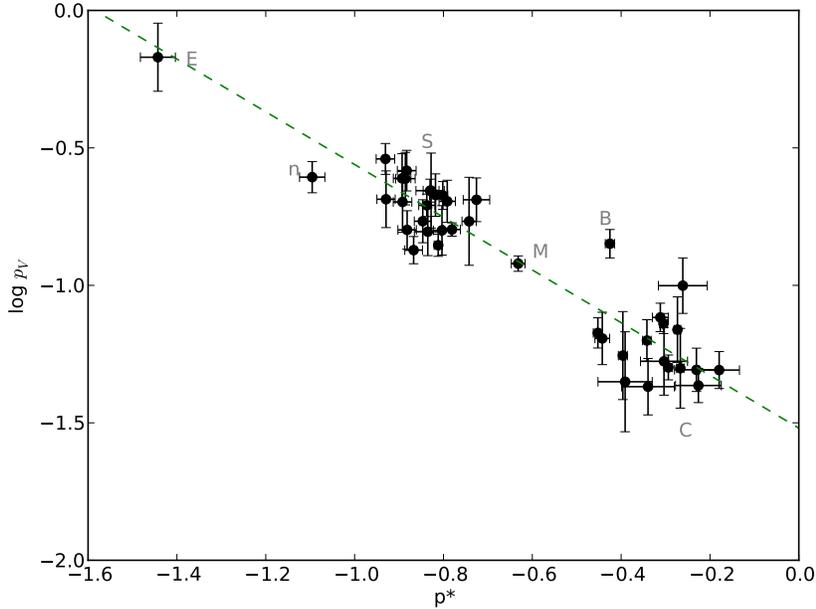} 
\protect\caption{The maximal variation of the polarimetric properties $p^\star$ as defined in the text compared to the measured albedo.  The best-fitting relationship projected from the three-dimensional fit is shown as the dashed line.  Text labels denote specific asteroids or groupings as discussed in the text.}
\label{fig.pstar}
\end{center}
\end{figure}

We can also project our fit of the polarimetric properties onto an
axis of maximal variation.  We define $p^\star$ as the quantity of
maximum polarimetric variation and find a best-fitting transform of:
\[p^\star = (0.79 \pm 0.02) \log h + (0.61 \pm 0.03) \log P_{min}\] 
Using our data we find a best-fitting relation between $p^\star$ and
albedo of:
\[\log p_V = (-1.58\pm0.09) + (-1.04\pm0.04) p^\star\]
We show $p^\star$ compared to albedo for all of the high-confidence
objects with both measured $h$ and $P_{min}$, along with this fit, in
Figure~\ref{fig.pstar}.  We label (64) Angelina as ``E'', (2) Pallas
as ``B'', and (132) Aethra as ``M'' following their Tholen taxonomic
classifications \citep{neesePDS}.  The clusters of objects with Tholen
S and C taxonomic classifications are labeled as such, and include
other objects within those taxonomic complexes (e.g. F- and D-types
are included in the C-complex) .  We indicate (71) Niobe with ``n'' in
this plot; though it has a Tholen class of S it is distinct enough
from the general cluster to warrant mention.  Additional polarimetric
and spectroscopic followup of this object will help determine why its
properties differ from the general S complex.  We focus on Tholen
classifications here as this taxonomic system shows the greatest
distinction in albedo between the different types
\citep{mainzer11tax}.

\begin{figure}[ht]
\begin{center}
\includegraphics[scale=0.5]{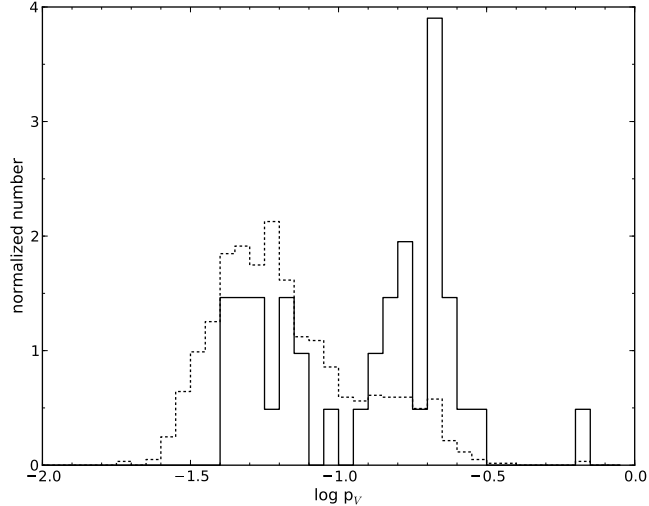} 
\protect\caption{Normalized albedo distribution of the high-confidence polarimetric sample (solid) and all MBAs larger than $30~$km that were detected during the fully cryogenic portion of the NEOWISE survey (dotted).}
\label{fig.pdist}
\end{center}
\end{figure}

In addition to the lack of the lowest albedo objects in the
polarimetric sample, we observe an over-representation of high albedo
objects compared to the distribution for all similarly sized MBAs.
Figure~\ref{fig.pdist} shows the distribution of albedos of all
objects in the sample with high-confidence polarimetric properties
used to derive the linear three dimensional fit (the smallest of which
is $D\sim35~$km), as well as all MBAs larger than $30~$km in diameter
that were observed by WISE.  The difference in these two distributions
can be traced to the optical selection bias in the acquisition and
measurement of the polarimetric properties of asteroids: very high
signal-to-noise levels are needed to reach the polarimetric
sensitivities that allow for accurate measurement of $P_{min}$ and
$h$.  Thus, even though at a given phase angle low albedo objects will
show larger degrees of polarization, the reduction in photons received
from these sources make these measurements less precise.  Focusing
future polarimetric surveys on low albedo asteroids will help make
this sample more representative of the true distribution of MBAs.

\section{Conclusions}

Using the newly available albedos from the WISE space telescope
thermal infrared survey data, we have fit the relationships between
albedo, the slope of the polarimetric-phase curve beyond the inversion
angle, and the maximum depth of the negative polarization trough.  We
restrict ourselves to objects with well characterized polarimetric
properties (i.e. relative errors $<20\%$).  Due to the selection of
the polarimetrically observed objects this results in our sample
consisting of only objects larger than $D>30~$km, with nearly
three-quarters having $D>50~$km.

We find that the function that best describes the albedo and
polarimetry is a three dimensional linear fit in $\log p_V$-$\log
h$-$\log P_{min}$ space.  Orthogonal distance regression allows us to
find the best fitting parameters while accounting for measurement
error on all parameters.  When the best fit line is projected to two
dimensions, we find the resultant fit parameters are all within $1.5$
sigma of those found by \citet{cellino99}.  We also define a new
polarimetric quantity $p^\star$ that describes the maximum variation
in polarimetric properties when compared with albedo.

We observe distinct separation of some taxonomic classes in $p^\star$
space.  In particular E-type, B-type, and some M-type asteroids are
far removed from the clumps that trace the more generic S- and
C-complex objects.  Asteroid (71) Niobe also holds a distinct location
in this space despite its S-type classification under the Tholen
system, and warrants further study.  We note that the principal
component (PC) analysis of Niobe from the Eight Color Asteroid Survey
indicates that it is on the edge of the S-complex \citep{tholenThesis}
and it has a PC4 in the bottom two percent of all asteroids in that
survey \citep[one of only 3 S-type or probable S-type objects with PC4 that
low][]{ecasPDS}.

Finally, despite the prevalence of low albedo asteroids seen
throughout the Main Belt \citep{masiero11} we find that they are
under-represented in the polarimetric sample.  Notably, roughly $10\%$
of MBAs have albedos $p_V<0.04$, but there are no objects in our
polarimetric sample with albedos this low.  We recommend that future
surveys focus on measuring polarization-phase curves for low albedo
asteroids to properly sample this population.

\section*{Acknowledgments}

The authors wish to thank referee Alberto Cellino for his helpful
review of this paper.  J.R.M. was supported by an appointment to the
NASA Postdoctoral Program at JPL, administered by Oak Ridge Associated
Universities through a contract with NASA.  This publication makes use
of data products from the Wide-field Infrared Survey Explorer, which
is a joint project of the University of California, Los Angeles, and
the Jet Propulsion Laboratory/California Institute of Technology,
funded by the National Aeronautics and Space Administration.  This
publication also makes use of data products from NEOWISE, which is a
project of the Jet Propulsion Laboratory/California Institute of
Technology, funded by the Planetary Science Division of the National
Aeronautics and Space Administration.  This research has made use of
the NASA/IPAC Infrared Science Archive, which is operated by the Jet
Propulsion Laboratory, California Institute of Technology, under
contract with the National Aeronautics and Space Administration.

\end{document}